\documentclass[amssymb,prc,twocolumn,superscriptaddress,nofootinbib]{revtex4-2}
\usepackage{graphicx}
\usepackage{amsmath}
\usepackage[T1]{fontenc}
\usepackage[utf8]{inputenc}

\begin{document}
	
\title[\empty]{Ab initio study of  \boldmath$Z(N) = 6$ 
magicity}

\author{H.~Li}
\affiliation{Institute of Modern Physics, Chinese Academy of Sciences, Lanzhou 730000, China}
\affiliation{School of Nuclear Science and Technology, Lanzhou University, Lanzhou 730000, China}
\affiliation{School of Nuclear Physics, University of Chinese Academy of Sciences, Beijing 100049, China}
\affiliation{CAS Key Laboratory of High Precision Nuclear Spectroscopy, Institute of Modern Physics, Chinese Academy of Sciences, Lanzhou 730000, China}

\author{H.~J.~Ong}
\affiliation{Institute of Modern Physics, Chinese Academy of Sciences, Lanzhou 730000, China}
\affiliation{School of Nuclear Physics, University of Chinese Academy of Sciences, Beijing 100049, China}
\affiliation{Joint Department for Nuclear Physics, Lanzhou University and Institute of Modern Physics, Chinese Academy of Sciences, Lanzhou 730000, China}
\affiliation{Research Center for Nuclear Physics, Osaka University, Ibaraki, Osaka 5670047, Japan}
\affiliation{RIKEN Nishina Center, Wako, Saitama 3510198,Japan}

\author{D.~Fang}
\affiliation{Institute of Modern Physics, Chinese Academy of Sciences, Lanzhou 730000, China}
\affiliation{School of Nuclear Physics, University of Chinese Academy of Sciences, Beijing 100049, China}

\author{I. A. Mazur}
\affiliation{Center for Exotic Nuclear Studies, Institute for Basic Science, Daejeon 34126, Republic of Korea}

\author{I. J. Shin}
\affiliation{Institute for Rare Isotope Science, Institute for Basic Science, Daejeon 34000, Republic of Korea}

\author{A.~M.~Shirokov}
\affiliation{Skobeltsyn Institute of Nuclear Physics, Lomonosov Moscow State University, Moscow 119991, Russia}

\author{J.~P.~Vary}
\affiliation{Department of Physics and Astronomy, Iowa State University, Ames, Iowa 50011, USA}

\author{P.~Yin}
\email[Corresponding author: ]{pengyin@iastate.edu}
\affiliation{College of Physics and Engineering, Henan University of Science and Technology, Luoyang 471023, China}
\affiliation{CAS Key Laboratory of High Precision Nuclear Spectroscopy, Institute of Modern Physics, Chinese Academy of Sciences, Lanzhou 730000, China}

\author{X.~Zhao}
\affiliation{Institute of Modern Physics, Chinese Academy of Sciences, Lanzhou 730000, China}
\affiliation{School of Nuclear Physics, University of Chinese Academy of Sciences, Beijing 100049, China}
\affiliation{CAS Key Laboratory of High Precision Nuclear Spectroscopy, Institute of Modern Physics, Chinese Academy of Sciences, Lanzhou 730000, China}

\author{W.~Zuo}
\affiliation{Institute of Modern Physics, Chinese Academy of Sciences, Lanzhou 730000, China}
\affiliation{School of Nuclear Physics, University of Chinese Academy of Sciences, Beijing 100049, China}
\affiliation{CAS Key Laboratory of High Precision Nuclear Spectroscopy, Institute of Modern Physics, Chinese Academy of Sciences, Lanzhou 730000, China}

\begin{abstract}
	The existence of   magic numbers   of protons and neutrons in nuclei is essential for understanding nuclear structure and fundamental nuclear forces. Over decades, researchers have conducted theoretical and experimental studies on the new magic number  { $Z(N) = 6$}, focusing on observables such as radii, binding energy, electromagnetic transition, and nucleon separation energies. We perform the {\it ab initio} no-core shell model calculations for the occupation numbers of  {the lowest} single particle states in  {the ground states of}
 {$Z(N)=6$ and $Z(N)=8$ isotopes (isotones)}. Our calculations do not support  {$Z(N)=6$} as a magic number over a span of atomic numbers. However, $^{14}$C and $^{14}$O exhibit the characteristics of double-magic nuclei.
\end{abstract}

\maketitle

The nuclear shell structure arises from the independent motion of nucleons in an average
 {mean-field}, serving as a valuable framework for comprehending nuclear structure and 
fundamental nuclear potential. The most notable characteristic of the shell structure is the presence of the so-called   magic numbers   of protons and neutrons, which are associated with enhanced stability. The occupation of nuclear shells results in the formation of nuclei with magic numbers. The introduction of the phenomenological strong nuclear force, which relies on the inherent spin and orbital angular momentum of a nucleon, along with the total angular momentum (orbital plus spin) coupling scheme~\cite{Mayer:1949pd,Haxel:1949fjd}, played a crucial role in fully explaining the magic numbers. This breakthrough earned Goeppert-Mayer and Jensen the Nobel Prize. 

Evidence from a multitude of experimental and theoretical studies have detailed the existence of nuclear shell structures. These findings have also revealed new magic numbers and the absence of previously recognized magic numbers  
in different regions of the periodic chart as well as appearance of the so-called local
magic numbers~\cite{Boboshin:2023ynd}.  A limited number of studies proposed non-traditional magicities 
like~$N = 14$~\cite{Stanoiu:2004nm,Brown:2005ds,Becheva:2006zz}, 
$N = 16$~\cite{Kanungo:2002jlq,Hoffman:2009zza,Tshoo:2012bi}, 
$Z = 16$~\cite{Togano:2012uj}, 
$N = 32$~\cite{Gade:2006dp,Wienholtz:2013nya,Rosenbusch:2015yma}, 
$N = 34$~\cite{Steppenbeck:2013mga}. 

In the realm of light nuclei, 
Kanungo {\it et al.}~\cite{Kanungo:2002jlq} studied in detail the sub-shell closure at $N = 6$ for neutron-rich isotopes
based on the analysis of separation energy systematics, beta decay Q-values, and the first excited states of nuclei. In addition, Otsuka {\it et al.}~\cite{Otsuka:2001nw} have examined the magicity of $N = 6$ in relation to the spin-isospin dependent component of the nucleon-nucleon interaction. Furthermore, studies utilizing the extension of the Bethe--Weizscker mass formula~\cite{Samanta:2002sp}, potential energy surfaces 
within the cluster-core model~\cite{Gupta:2006sd} and relativistic mean-field theory~\cite{Kumawat:2018ghm} have demonstrated that $N = 6$ and $Z = 6$ exhibit traits 
similar to those of  {the} shell closures. A persistent $Z=6$ magicity in $^{13-20}$C was proposed ~\cite{Tran:2017gxv} based on systematic analyses of radii, electromagnetic transition rates, and nuclear masses in
carbon and neighboring isotopes using published data, their new experimental data as well as results of shell model and \textit{ab initio}
coupled-cluster calculations with modern chiral effective field theory inter-nucleon interactions
obtained by themself and those from
Ref.~\cite{Forssen:2011dr}.

\begin{figure*}
	\includegraphics[width=\textwidth]{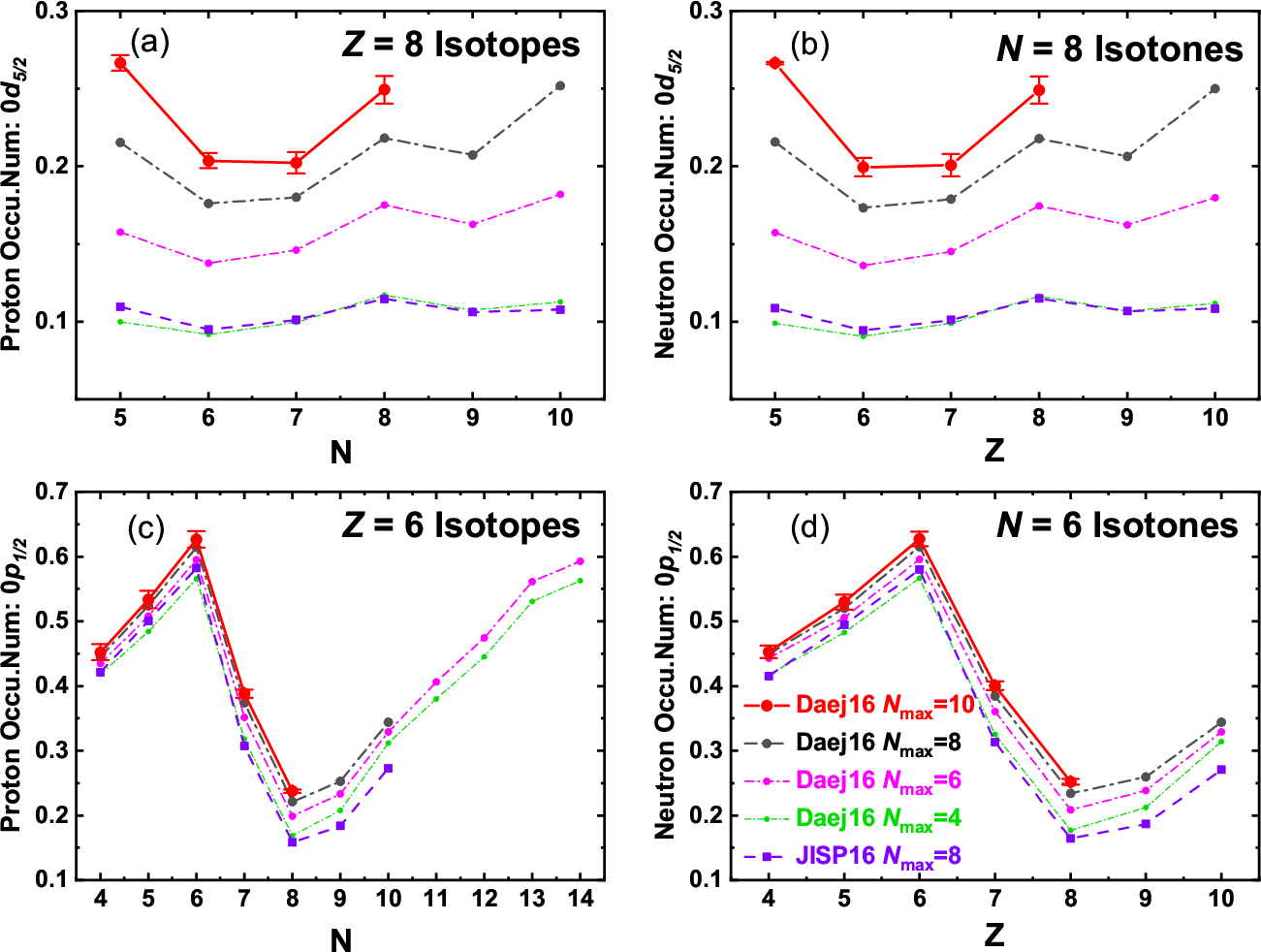}
	\caption{\label{fig:Occu.Num} Ground state occupation numbers  of
		(a)~protons in $0d_{5/2}$ state in oxygen isotopes, (b)~neutrons in $0d_{5/2}$ state in  $N = 8$ isotones, (c)~protons in $0p_{1/2}$ state in carbon isotopes, and (d)~neutrons in  $0p_{1/2}$ state in $N = 6$ isotones. 
		{The NCSM calculations performed} in a harmonic oscillator basis using the Daejeon16 (Daej16)  {$NN$ interaction
			with $N_{\max}$ ranging from 4} to 10  {and $\hbar\Omega = 17.5$~MeV.
			The results obtained with the JISP16 $NN$ interaction} at $N_{\max}= 8$ are shown for comparison. 
	}
\end{figure*}

  {In this contribution, we examine the ${Z=6}$ magicity of carbon isotopes
and  magicity of their mirror  $N=6$ isotones by {\it ab initio} no-core shell model
(NCSM)~\cite{Barrett:2013nh,Navratil:2000gs,Navratil:2000ww} calculations of occupation numbers 
in the lowest oscillator single-particle states of these nuclei. Although the oscillator states are not as preferred as natural orbitals or Hartree-Fock orbitals, we find them sufficient for our purposes. In fact, we establish that the~$0s_{1/2} + 0p_{3/2}$ occupancies are near their maxima and we compare the $0p_{1/2}$ occupancies with the~$0d_{5/2}$ occupancies in the nuclei with the well-established magic numbers~$Z=8$ and~$N=8$.}
Our NCSM results are obtained with the Daejeon16~\cite{Shirokov:2016ead} 
$NN$ interaction. {The} Daejeon16 interaction is based on the Entem--Machleidt N$^3$LO chiral effective field theory interaction~\cite{Entem:2003ft}, softened via a similarity renormalization group transformation~\cite{Bogner:2006pc}  to provide 
 {a faster} convergence, and then adjusted via phase-shift equivalent  {transformations
(PETs) to provide a good description of} nuclei with $A\le16$  {without the use of $3N$ forces
whose effects are mimiced by the PET modification of the $NN$ interaction's
off-shell properties.} Using the MFDn code~\cite{MARIS:2010, Aktulga:2013}, we diagonalize the 
Hamiltonian of the  {nuclear} system in a many-body harmonic oscillator basis  which is characterized by the basis energy scale~$\hbar\Omega$ and the basis truncation parameter $N_{\max}$, the maximum number of oscillator excitation quanta allowed in the many-body space relative to the lowest Pauli-allowed configuration.

The NCSM calculations  {with} the Daejeon16 $NN$ interaction
 {are performed wherever possible}
 in the model  {spaces $N_{\max}=4$, 6, 8
 and~10. However, due to a fast growth 
 of the model space dimension and hence the computational cost,
we first drop calculations
with~$N_{\max}=10$ as the number of nucleons~$A$ increases,  and next additionally 
with~$N_{\max}=8$ after a further increase of~$A$.} 
We  {present 
in Fig.~\ref{fig:Occu.Num} the results 
for occupation numbers in low-lying harmonic oscillator single-particle states
obtained with oscillator energy}~$\hbar\Omega = 17.5$~MeV.  {This~$\hbar\Omega$ value corresponds to}  
the minimum of  {the ground state} energy obtained  {with} the largest $N_{\max}$  {for nearly all nuclei
discussed here where we have used a grid with 2.5 MeV increments in $\hbar\Omega$}.  {The only exception is} $^{13}$O; however, the $\hbar\Omega$ dependence of
the occupation numbers is weak and in $^{13}$O they differ by less than 2\%
at~$\hbar\Omega = 15$~MeV corresponding to the
ground state energy minimum in this nucleus. To visualize the $\hbar\Omega$ dependence,
for the results at the largest model space we present   {`error bars'} where the lowest (highest) point 
indicates the minimal (maximal) occupation number value in the range 
from~$\hbar\Omega = 15$ to 20~MeV.

In the {``naive shell model''} with non-interacting nucleons moving in the mean-field, 
the proton or neutron orbitals above the closed shells associated with respective 
magic~$Z$ or~$N$ numbers, are completely unoccupied. Conversely, orbitals at or below closed shells are occupied at their maximum occupation number of $2j + 1$. 

In the case of the NCSM calculations with Daejeon16 $NN$ interaction, the occupation numbers of protons and neutrons 
in the~$0d_{5/2}$ 
single-particle state, the first single-particle state above the $Z=8$ and $N=8$ shell closures 
in respectively oxygen   ($Z=8$) isotopes and $N=8$ isotones, are ranging from 
approximately~0.1 in the smallest presented model space with~$N_{\max}=4$
to approximately~0.27 in the largest presented model space with~$N_{\max}=10$
as is seen in Figs.~\ref{fig:Occu.Num}(a,b). {Based on the results of the~$0d_{5/2}$  occupations in $Z=8$ isotopes and 
$N=8$ isotones, we suggest 
to set the proton (neutron) occupation number value of around 0.27 for the~$0p_{1/2}$ orbital at $N_{\max}=10$
 as an approximate border of magicity in $Z(N)=6$ isotopes (isotones).}

 {Supposing that $Z=6$ is a ``good'' magic number in carbon isotopes, we should expect
that the $0p_{3/2}$ orbital is nearly completely occupied while the occupation numbers
in the $0p_{1/2}$ orbital are similar to those of protons and neutrons in
the $0d_{5/2}$ single-particle states in the oxygen isotopes and $N=8$ isotones. The $0p_{1/2}$}
 proton occupation numbers in 
{$^{10\mbox{-}20}$C} isotopes
 {are presented} in Fig.~\ref{fig:Occu.Num}(c). There is an increasing trend
 {with~$A$} observed between {$^{10}$C} and $^{12}$C  {where the $0p_{1/2}$ occupation numbers are approximately
0.4--0.6, i.\,e., much larger than is expected for the closed $0p_{3/2}$ subshell. With~$A$ increasing further, we observe a sharp decline of the $0p_{1/2}$ occupation number. Especially, the one in $^{14}$C falls into the range of 0.2--0.25, which is what we expect for ``good'' magic numbers. Beyond $^{14}$C we see a gradual rise of the $0p_{1/2}$ occupation number, and in $^{20}$C it attains approximately the same value as in $^{12}$C.
}

 {From our results it follows that
the $^{14}$C nucleus 
with $Z=6$ and $N=8$
does indeed exhibit features of a double-magic nucleus, supporting the same proposition based on
the analysis of various data in Ref.~\cite{Tran:2017gxv}. However, the $Z=6$ magicity is
weakened in carbon isotopes with $N<8$ and $N>9$.
From the analysis of the occupation numbers in the $0p_{1/2}$ orbital in carbon isotopes, we 
can conclude that $Z=6$ is a {\it local magic number} which exhibits marked magic features 
when the number of neutrons lies in the interval $8\lesssim N \lesssim9$.
}

Regarding the carbon isotopes, we note that $^{12}$C is well-known to have a three-alpha structure associated with the deformed oblate shape which is manifested in a rotational band including the ground and the first $2^{+}$ states of $^{12}$C as members~\cite{Otsuka:2022bcf}. The double-magic nuclei are known to maintain a spherical shape, and a switch from deformation to sphericity results in a strong drop-off of $0p_{1/2}$ proton occupation between $^{12}$C and $^{14}$C as seen in Fig.~\ref{fig:Occu.Num}(c). As the number of neutrons increases in the $N>8$ carbon isotopes, the deformation is slowly increasing which results in disappearance of magicity.

Figure~\ref{fig:Occu.Num}(d) displays the neutron occupation numbers 
 {in the $0p_{1/2}$ orbital} for $N = 6$ isotones from 
{$Z = 4$} to $Z = 10$.  {We calculated fewer $N = 6$ isotones than carbon isotopes since
proton-excess nuclei with  $N = 6$  and $Z > 8$ are particle-unstable.}
 The results  {for} $N = 6$ isotones exhibit concordance with carbon isotopes: 
 {the growths of the $0p_{1/2}$ neutron occupation numbers from {$^{10}$Be} to $^{12}$C, 
follows by a sharp drop towards $^{14}$O, which appears to exhibit the double-magic features,
and a gradual increase in $^{15}$F and $^{16}$Ne. Thus we conclude that $N=6$ is
also a {\it local magic number}
 in the interval  $8\lesssim Z \lesssim9$.
}

{ {To ensure the independence on the
$NN$ interaction of our qualitative conclusions about the {\it local magicity}
of $Z=6$ and $N=6$, we performed also the NCSM calculations with the  JISP16 
$NN$ interaction~\cite{Shirokov:2005bk}. The origin of the}
JISP16 interaction  {is very different from that of Daejeon16: it
was} initially developed from  {the} $NN$  {scattering} data using inverse scattering
techniques,  {and then, as with Daejeon16,}
adjusted  {by PETs to $A\le16$ nuclei to avoid the need in the $3N$ forces.}

The JISP16 results for the occupation numbers  were obtained with $N_{\max} = 8$.  
Although the minimum of binding energy obtained with JISP16 usually appear at higher 
$\hbar\Omega$ values, we used also $\hbar\Omega = 17.5$~MeV in the JISP16 calculations. 
The differences between the JISP16 results presented in Fig.~\ref{fig:Occu.Num} (obtained at $\hbar\Omega = 17.5$~MeV) and
 occupation numbers  calculated at  the  $\hbar\Omega$ values corresponding to the minima 
of the ground state energies does not exceed~10\%. 
The results obtained from the JISP16 calculation exhibit modest variations compared to those from Daejeon16, however, the overall pattern of change remains constant.

 {The occupation numbers obtained with JISP16 clearly follow the same trends. The
JISP16 occupations are somewhat smaller than those supported by the Daejeon16.  For
example, the proton $0p_{1/2}$ occupation number in $^{14}$C is about~0.16, which is approximately 30\% smaller than the respective Daejeon16 result from the same model space.
Note, however, that the JISP16 predicts also much smaller 
occupation numbers in the $0d_{5/2}$ orbital just above proton
$Z=8$ and neutron $N=8$ shell closures than the Daejeon16 $NN$ interaction. Therefore
we again obtain approximately the same occupation numbers for the proton $0p_{1/2}$
orbital and neutron $0d_{5/2}$ orbital in $^{14}$C. Generally, the JISP16 calculations
support our conclusion that the $^{14}$C and $^{14}$O nuclei have the double-magic features.
However, both $Z=6$ and $N=6$ are 
 {\it local magic numbers}, whose magicity reveals markedly
only in the intervals of neutron numbers $8\lesssim N \lesssim9$ and proton numbers $8\lesssim Z \lesssim9$,
respectively.
}
 
 {Concluding, we have examined in the {\it ab initio} NCSM calculations
 the proton (neutron) occupation numbers in the  $0p_{1/2}$
orbital in the carbon isotopes ($N=6$  isotones) 
and in the $0d_{5/2}$ orbital in oxygen isotopes ($N=8$ isotones). We have checked that the occupancies of orbitals above $0p_{1/2}$ are significantly smaller than the value of $0p_{1/2}$ in $Z(N)=6$ isotopes (isotones). Correspondingly, we have also found nearly complete occupation of the $0s_{1/2} + 0p_{3/2}$ orbitals which provides a foundation for examining the magicity of $Z(N) = 6$. 
Our analysis supports the proposition of Ref.~\cite{Tran:2017gxv} that $^{14}$C and $^{14}$O
are double-magic nuclei,
and suggests that $Z=6$ and $N=6$ are  {\it local magic numbers} 
which weaken the magicity when respectively neutron numbers are approximately beyond the interval
$8\lesssim N \lesssim9$ and proton numbers are approximately beyond the interval
$8\lesssim Z \lesssim9$. It will be interesting to perform spectroscopic studies using proton- or neutron-transfer and/or knockout reactions on the relevant $Z(N)=6$ isotopes (isotones) to investigate the structural evolution. 
}

\mbox{}

{\it Acknowledgements.}
This work is partially supported by the National Natural Science Foundation of China (Grant Nos.~12175280, 12250610193, 12375143, 11975282, 11705240, 11435014), the Natural Science Foundation of Gansu Province (Grant Nos.~20JR10RA067,~23JRRA675), the Chinese Academy of Sciences ``Light of West China'' Program, the Key Research Program of the Chinese Academy of Sciences (Grant Nos.~ZDB-SLY-7020, XDPB15), the Foundation for Key Talents of Gansu Province by the Central Funds Guiding the Local Science and Technology Development of Gansu Province (Grant No.~22ZY1QA006), the International Partnership Program of the Chinese Academy of Sciences (Grant No.~016GJHZ2022103FN), the National Key R$\&$D Program of China (Grant No.~2023YFA1606903), the Strategic Priority Research Program of the Chinese Academy of Sciences (Grant No.~XDB34000000), 
the Gansu International Collaboration and Talents Recruitment Base of Particle Physics (2023–2027), the US Department of Energy (Grant No.~DE-SC0023692), the National Research Foundation of Korea (2013M7A1A1075764). The Chinese Academy of Sciences President’s International Fellowship Initiative (Grant Nos.~2023VMA0013) provides financing for A.~M.~Shirokov's trip to China to participate in this work. I.~A.~Mazur is supported by the Institute for Basic Science (IBS-R031-D1).  A portion of the computational resources were provided by Gansu Computing Center, Sugon Computing Center in Xi'an. Computational resources including technical support were also partly provided by the National Supercomputing Center of Korea (KSC-2024-CHA- 0001).

\end{document}